\documentclass[12pt]{article}
\usepackage{amsmath,amssymb,amsthm,amsxtra,overpic,bbm,bm,epsfig,ulem}
\usepackage{color}
\usepackage{cite}

\usepackage{hyperref}
\usepackage{url}

\def\thefootnote{\fnsymbol{footnote}}

\begin{document}

\vspace{0.2cm}

\begin{center}
{\large\bf A Pythagoras-like theorem for CP violation in neutrino oscillations}
\end{center}

\vspace{0.2cm}

\begin{center}
{\bf Shu Luo}$^1$
\footnote{E-mail: luoshu@xmu.edu.cn}
and
{\bf Zhi-zhong Xing$^{2,3}$}
\footnote{E-mail: xingzz@ihep.ac.cn}
\\
{\small $^{1}$Department of Astronomy, Xiamen University, Fujian 361005, China} \\
{\small $^{2}$Institute of High Energy Physics and School of Physical Sciences, \\
University of Chinese Academy of Sciences, Beijing 100049, China \\
$^{3}$Center of High Energy Physics, Peking University, Beijing 100871, China}
\end{center}

\vspace{2cm}
\begin{abstract}
The probabilities of $\nu^{}_\mu \to \nu^{}_e$ and $\overline{\nu}^{}_\mu \to
\overline{\nu}^{}_e$ oscillations in vacuum are determined by the CP-conserving
flavor mixing factors ${\cal R}^{}_{ij} \equiv {\rm Re} (U^{}_{\mu i} U^{}_{e j}
U^*_{\mu j} U^*_{e i})$ and the universal Jarlskog invariant of CP violation
${\cal J}^{}_\nu \equiv (-1)^{i+j} \hspace{0.05cm} {\rm Im} (U^{}_{\mu i} U^{}_{e j}
U^*_{\mu j} U^*_{e i})$ (for $i, j = 1, 2, 3$ and $i < j$), where $U$ is the
$3\times 3$ Pontecorvo-Maki-Nakagawa-Sakata neutrino mixing matrix. We show that
${\cal J}^{2}_\nu = {\cal R}^{}_{12} {\cal R}^{}_{13} +
{\cal R}^{}_{12} {\cal R}^{}_{23} + {\cal R}^{}_{13} {\cal R}^{}_{23}$
holds as a natural consequence of the unitarity of $U$. This Pythagoras-like
relation may provide a novel cross-check of the result of ${\cal J}^{}_\nu$
that will be directly measured in the next-generation long-baseline neutrino
oscillation experiments. Indirect non-unitarity effects and terrestrial matter
effects on ${\cal J}^{}_\nu$ and ${\cal R}^{}_{ij}$ are also
discussed.
\end{abstract}

\newpage

\def\thefootnote{\arabic{footnote}}
\setcounter{footnote}{0}
\setcounter{figure}{0}

\section{Introduction}

The experimental discoveries of atmospheric, solar,
reactor and accelerator neutrino oscillations have
constituted the most remarkable evidence for new physics beyond the
standard model (SM) of particle physics~\cite{ParticleDataGroup:2022pth}.
The relevant data can be well described by the $3\times 3$
Pontecorvo-Maki-Nakagawa-Sakata (PMNS) lepton flavor mixing matrix
$U$~\cite{Pontecorvo:1957cp,Maki:1962mu,Pontecorvo:1967fh}, together with
the neutrino mass-squared differences $\Delta^{}_{ij} \equiv m^2_i - m^2_j$
(for $i, j = 1, 2, 3$ and $i \neq j$). Although the unitarity of $U$ has
been tested at the level of
${\cal O}\left(\lesssim 10^{-2}\right)$~\cite{Antusch:2006vwa,
Antusch:2009gn,Blennow:2016jkn,Hu:2020oba,Wang:2021rsi,Blennow:2023mqx},
whether it is exactly unitary or not actually depends on the theoretical
origin of tiny neutrino masses
\footnote{For example, the $3\times 3$ PMNS matrix $U$ is exactly unitary
in the type-II seesaw mechanism~\cite{Konetschny:1977bn,Magg:1980ut,
Schechter:1980gr,Cheng:1980qt,Lazarides:1980nt,Mohapatra:1980yp},
but it is non-unitary to some degree either in the type-I seesaw
mechanism~\cite{Minkowski:1977sc,Yanagida:1979as,GellMann:1980vs,
Glashow:1979nm,Mohapatra:1979ia} or in the type-III seesaw
scenario~\cite{Foot:1988aq}.}.
That is why the next-generation neutrino oscillation experiments aim
to convincingly establish leptonic CP violation and accurately measure
all the flavor mixing parameters that are accessible in both the {\it appearance}
(i.e., $\nu^{}_\alpha \to \nu^{}_\beta$ and $\overline{\nu}^{}_\alpha
\to \overline{\nu}^{}_\beta$ with $\beta \neq \alpha$) and
{\it disappearance} (i.e., $\nu^{}_\alpha \to \nu^{}_\alpha$ and
$\overline{\nu}^{}_\alpha \to \overline{\nu}^{}_\alpha$ for $\alpha
= e, \mu, \tau$) oscillation channels. Among them, $\nu^{}_\mu
\to \nu^{}_e$ and $\overline{\nu}^{}_\mu \to \overline{\nu}^{}_e$
oscillations are the main focus for a realistic experimental exploration
of CP violation in the neutrino sector. A preliminary but encouraging
evidence for CP violation in these two channels has recently be reported
by the T2K Collaboration at the $2\sigma$ level~\cite{T2K:2019bcf,T2K:2023smv}.

Given unitarity of the $3\times 3$ PMNS matrix $U$, the probabilities of
$\nu^{}_\mu \to \nu^{}_e$ and $\overline{\nu}^{}_\mu \to \overline{\nu}^{}_e$
oscillations in vacuum are well known~\cite{ParticleDataGroup:2022pth} and
can be compactly expressed as
\begin{eqnarray}
P(\nu^{}_\mu \to \nu^{}_e) \hspace{-0.2cm} & = & \hspace{-0.2cm}
- 4 \sum_{i<j} \left({\cal R}^{}_{ij}
\sin^2 \frac{\Delta^{}_{ji} L}{4 E} \right)
- 8 {\cal J}^{}_\nu \prod_{i<j} \sin \frac{\Delta^{}_{ji} L}{4 E} \; ,
\nonumber \\
P(\overline{\nu}^{}_\mu \to \overline{\nu}^{}_e)
\hspace{-0.2cm} & = & \hspace{-0.2cm}
- 4 \sum_{i<j} \left({\cal R}^{}_{ij}
\sin^2 \frac{\Delta^{}_{ji} L}{4 E} \right)
+ 8 {\cal J}^{}_\nu \prod_{i<j} \sin \frac{\Delta^{}_{ji} L}{4 E} \; ,
\hspace{0.4cm}
\label{1}
\end{eqnarray}
where $L$ is the baseline length, $E$ denotes the neutrino beam energy,
${\cal R}^{}_{ij} \equiv {\rm Re}
(U^{}_{\mu i} U^{}_{e j} U^*_{\mu j} U^*_{e i})$ are the
CP-conserving flavor mixing factors, and ${\cal J}^{}_\nu =
\left(-1\right)^{i+j} \hspace{0.05cm} {\rm Im} (U^{}_{\mu i}
U^{}_{e j} U^*_{\mu j} U^*_{e i})$ is the Jarlskog invariant
of leptonic CP violation (for $i, j = 1, 2, 3$ and
$i < j$)~\cite{Jarlskog:1985ht,Wu:1985ea,Cheng:1986in}. It is expected
that ${\cal R}^{}_{ij}$ and ${\cal J}^{}_\nu$ will all be determined
from a precision measurement of the dependence of $P(\nu^{}_\mu \to \nu^{}_e)$
or $P(\overline{\nu}^{}_\mu \to \overline{\nu}^{}_e)$ on $L/E$, from
which one may numerically examine whether there is a kind of correlation
between ${\cal J}^{}_\nu$ and ${\cal R}^{}_{ij}$.

In this short note we are going to show that the universal Jarlskog
invariant ${\cal J}^{}_\nu$ is actually related to the three
CP-conserving quantities ${\cal R}^{}_{12}$, ${\cal R}^{}_{13}$
and ${\cal R}^{}_{23}$ in a remarkably simple and interesting way:
${\cal J}^{2}_\nu = {\cal R}^{}_{12} {\cal R}^{}_{13} +
{\cal R}^{}_{12} {\cal R}^{}_{23} + {\cal R}^{}_{13}
{\cal R}^{}_{23}$, which will subsequently be referred to as a
Pythagoras-like theorem for CP violation in neutrino oscillations.
Such a novel relation allows us to {\it calculate} the size of
${\cal J}^{}_\nu$ from the experimental values of ${\cal R}^{}_{ij}$,
and thus it can provide a very useful cross-check of the result of
${\cal J}^{}_\nu$ that will be {\it directly measured} in the next-generation
$\nu^{}_\mu \to \nu^{}_e$ and $\overline{\nu}^{}_\mu \to \overline{\nu}^{}_e$
oscillation experiments (e.g., Hyper-K~\cite{Hyper-KamiokandeProto-:2015xww},
T2HKK~\cite{Hyper-Kamiokande:2016srs} and DUNE~\cite{DUNE:2020jqi}).
Since the above correlation between ${\cal J}^{}_\nu$ and ${\cal R}^{}_{ij}$ is
a natural consequence of the unitarity of the $3\times 3$ PMNS matrix $U$, an
experimental test of its validity is certainly important in the
precision measurement era of neutrino physics.

Note that $U$ will not be exactly unitary in the well-motivated seesaw
mechanism of type-I~\cite{Minkowski:1977sc,Yanagida:1979as,GellMann:1980vs,
Glashow:1979nm,Mohapatra:1979ia} (or type-III~\cite{Foot:1988aq})
due to the slight active-sterile flavor
mixing, although the relevant sterile (heavy) Majorana neutrinos are
kinematically forbidden to participate in the flavor oscillations of three
active (light) neutrinos. In this {\it indirect} unitarity violation
case, the formulas of $P(\nu^{}_\mu \to \nu^{}_e)$ and
$P(\overline{\nu}^{}_\mu \to \overline{\nu}^{}_e)$ in Eq.~(\ref{1}) have
to be modified to some extent. We are going to examine such non-unitarity
effects on ${\cal J}^{}_\nu$ and ${\cal R}^{}_{ij}$. Possible
terrestrial matter effects on ${\cal J}^{}_\nu$ and ${\cal R}^{}_{ij}$
in a realistic long-baseline experiment will also be discussed.

\section{The theorem in vacuum}

\begin{figure}[t]
\begin{center}
\vspace{-0.35cm}
\includegraphics[width=17.2cm]{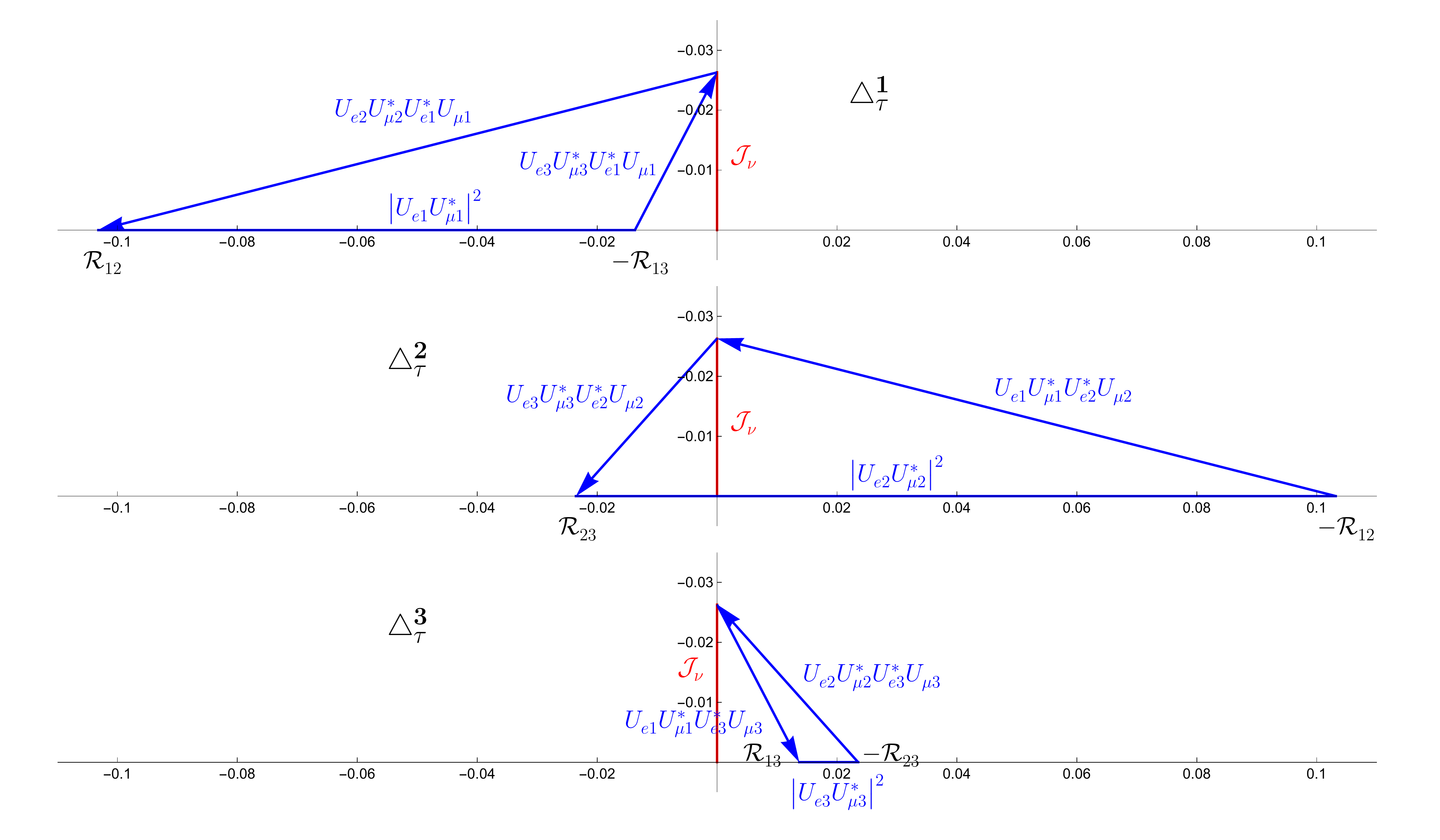}
\vspace{-1.2cm}
\caption{A numerical illustration of the three rescaled unitarity triangles
originating from $\triangle^{}_{\tau}$ in the complex plane for the
{\bf normal} neutrino mass ordering, where the best-fit values of the
relevant flavor mixing and CP-violating
parameters~\cite{Gonzalez-Garcia:2021dve,Nufit} have been
input.}
\label{fig1}
\end{center}
\end{figure}

\begin{figure}[t]
\begin{center}
\vspace{-0.3cm}
\includegraphics[width=17.2cm]{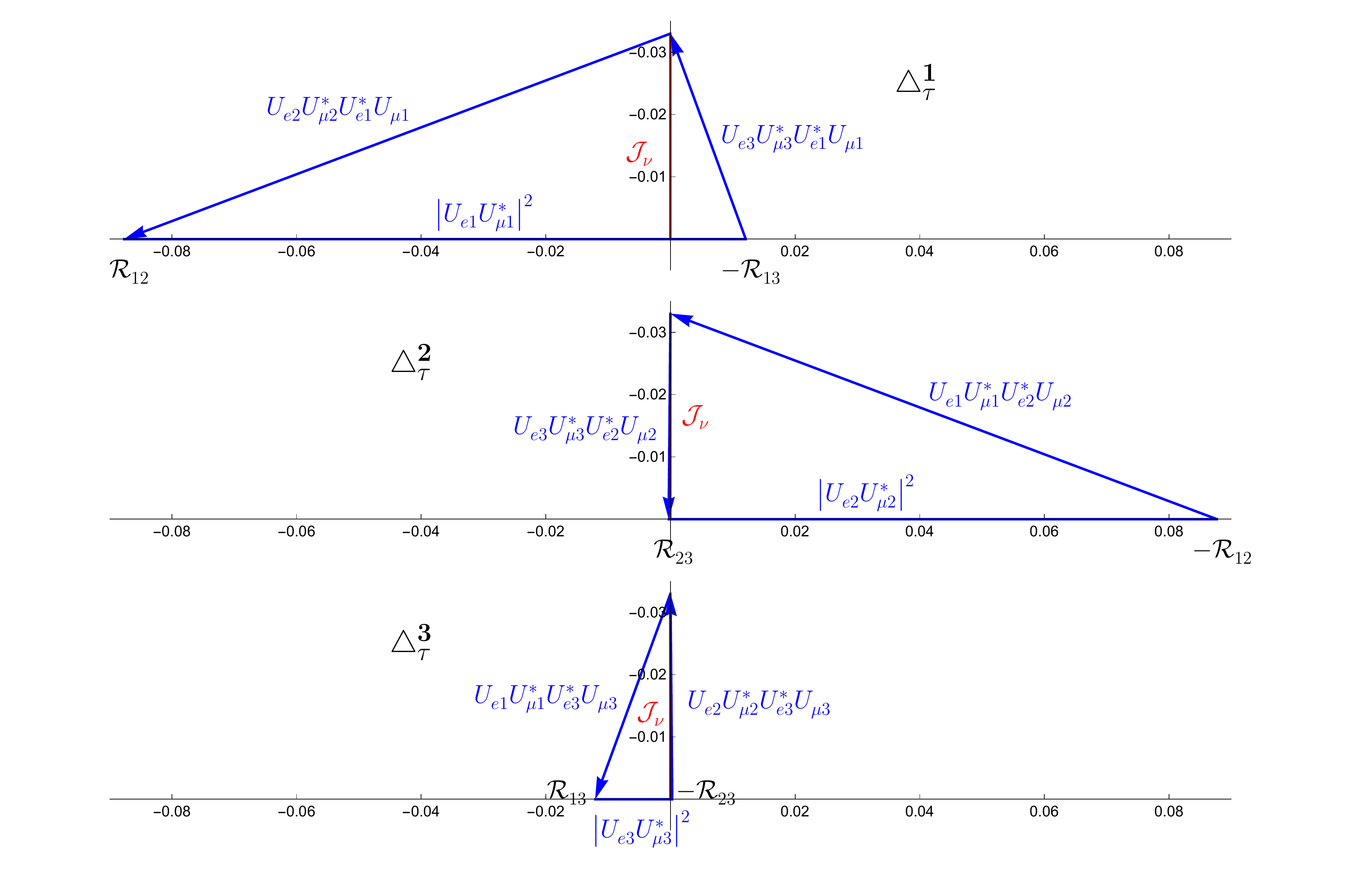}
\vspace{-1.2cm}
\caption{A numerical illustration of the three rescaled unitarity triangles
originating from $\triangle^{}_{\tau}$ in the complex plane for the
{\bf inverted} neutrino mass ordering, where the best-fit values of the
relevant flavor mixing and CP-violating
parameters~\cite{Gonzalez-Garcia:2021dve,Nufit} have been
input.}
\label{fig2}
\end{center}
\end{figure}

Let us begin with the orthogonality relation of the $3\times 3$
PMNS lepton flavor mixing matrix $U$,
\begin{eqnarray}
U^{}_{e 1} U^*_{\mu 1} + U^{}_{e 2} U^*_{\mu 2} +
U^{}_{e 3} U^*_{\mu 3} = 0 \; ,
\label{2}
\end{eqnarray}
which defines a triangle in the complex plane. This leptonic unitarity
triangle is referred to as $\triangle^{}_\tau$~\cite{Fritzsch:1999ee},
and it is closely associated with
$\nu^{}_\mu \to \nu^{}_e$ and $\overline{\nu}^{}_\mu \to
\overline{\nu}^{}_e$ oscillations. There are three straightforward ways
to rescale $\triangle^{}_\tau$, leading to three fully rephasing-invariant
triangles $\triangle^{\bf n}_\tau$ (for ${\bf n} = {\bf 1}, {\bf 2},
{\bf 3}$) whose real and positive sides are defined as
\begin{eqnarray}
\triangle^{\bf 1}_\tau : ~\quad {\cal S}^{2}_1
\hspace{-0.2cm} & \equiv & \hspace{-0.2cm}
\left|U^{}_{e 1} U^*_{\mu 1}\right|^2 = - U^{}_{\mu 1} U^{}_{e 2}
U^*_{\mu 2} U^*_{e 1} - U^{}_{\mu 1} U^{}_{e 3}
U^*_{\mu 3} U^*_{e 1} \; ,
\nonumber \\
\triangle^{\bf 2}_\tau : ~\quad {\cal S}^{2}_2
\hspace{-0.2cm} & \equiv & \hspace{-0.2cm}
\left|U^{}_{e 2} U^*_{\mu 2}\right|^2 = - U^{}_{\mu 2} U^{}_{e 3}
U^*_{\mu 3} U^*_{e 2} - U^{}_{\mu 2} U^{}_{e 1}
U^*_{\mu 1} U^*_{e 2} \; ,
\nonumber \\
\triangle^{\bf 3}_\tau : ~\quad {\cal S}^{2}_3
\hspace{-0.2cm} & \equiv & \hspace{-0.2cm}
\left|U^{}_{e 3} U^*_{\mu 3}\right|^2 = - U^{}_{\mu 3} U^{}_{e 1}
U^*_{\mu 1} U^*_{e 3} - U^{}_{\mu 3} U^{}_{e 2}
U^*_{\mu 2} U^*_{e 3} \; , \hspace{0.5cm}
\label{3}
\end{eqnarray}
respectively. In Fig.~\ref{fig1} (or Fig.~\ref{fig2}) we have numerically
illustrated the sizes and shapes of $\triangle^{\bf n}_\tau$ by using
the best-fit values $\theta^{}_{12} = 33.41^\circ$ (or $33.41^\circ$),
$\theta^{}_{13} = 8.58^\circ$ (or $8.57^\circ$),
$\theta^{}_{23} = 42.2^\circ$ (or $49.0^\circ$) and $\delta^{}_\nu = 232^\circ$
(or $276^\circ$) in the standard parametrization of $U$ corresponding to
the normal (or inverted) neutrino mass ordering,
as obtained from a recent global analysis of current neutrino oscillation
data~\cite{Gonzalez-Garcia:2021dve,Nufit}
\footnote{The corresponding best-fit values of the two independent
neutrino mass-squared differences are found to be
$\Delta^{}_{21} = 7.41 \times 10^{-5} ~{\rm eV}^2$ (or
$7.41 \times 10^{-5}~{\rm eV}^2$) and $\Delta^{}_{31} = +2.507 \times
10^{-3}~{\rm eV}^2$ (or $-2.412 \times 10^{-3}~{\rm eV}^2$) for the
normal (or inverted) neutrino mass spectrum in this global analysis.}.
As a result, we arrive at
\begin{eqnarray}
{\cal J}^{}_\nu = \frac{1}{8} \sin 2\theta^{}_{12}
\sin 2\theta^{}_{13} \sin 2\theta^{}_{23} \cos\theta^{}_{13}
\sin\delta^{}_\nu \simeq
\left\{\begin{array}{l} \hspace{-0.2cm} -2.63 \times 10^{-2} \\
\hspace{-0.2cm} -3.30 \times 10^{-2}
\end{array}
\right.
\label{4}
\end{eqnarray}
in the normal (upper) and inverted (lower) mass ordering cases, respectively.
One can see that the heights of triangles $\triangle^{\bf 1}_\tau$,
$\triangle^{\bf 2}_\tau$ and $\triangle^{\bf 3}_\tau$ corresponding to the
respective bases $|U^{}_{e 1} U^*_{\mu 1}|^2$, $|U^{}_{e 2} U^*_{\mu 2}|^2$
and $|U^{}_{e 3} U^*_{\mu 3}|^2$ are all equal to ${\cal J}^{}_\nu$ as
shown in Fig.~\ref{fig1} or Fig.~\ref{fig2},
a salient feature of the unitarity of $U$ which is similar to the case of the
rescaled unitarity triangles in the quark sector~\cite{Xing:2012zv}.
Moreover, we find that ${\cal R}^{}_{12}$ and ${\cal R}^{}_{23}$ are
negative and ${\cal R}^{}_{13}$ is positive in the normal mass ordering
case, while ${\cal R}^{}_{12}$, ${\cal R}^{}_{13}$ and ${\cal R}^{}_{23}$ are
all negative (and ${\cal R}^{}_{23}$ is close to zero) in the inverted mass
ordering case. This observation is certainly subject to the input best-fit values of
$\theta^{}_{12}$, $\theta^{}_{13}$, $\theta^{}_{23}$ and $\delta^{}_\nu$, which are
expected to more or less change in the near future,
but it can at least give us a ball-park feeling of the geometry of triangles
$\triangle^{\bf n}_\tau$ in the complex plane.

Note that triangle $\triangle^{\bf n}_\tau$ is apparently associated with two
right triangles which share the common side characterized by ${\cal J}^{}_\nu$
(for ${\bf n} = {\bf 1}, {\bf 2}, {\bf 3}$), as shown in Fig.~\ref{fig1} or
Fig.~\ref{fig2}. Focusing on the right triangle formed by its catheti
${\cal J}^{}_\nu$ and ${\cal R}^{}_{12}$ (in $\triangle^{\bf 1}_\tau$),
${\cal R}^{}_{23}$ (in $\triangle^{\bf 2}_\tau$) or ${\cal R}^{}_{13}$
(in $\triangle^{\bf 3}_\tau$), we follow the famous Pythagorean theorem and
obtain the interesting relations
\begin{eqnarray}
{\cal S}^{2}_1 {\cal S}^{2}_2 \hspace{-0.2cm} & = & \hspace{-0.2cm}
{\cal R}^{2}_{12} + {\cal J}^{2}_\nu \; ,
\nonumber \\
{\cal S}^{2}_1 {\cal S}^{2}_3 \hspace{-0.2cm} & = & \hspace{-0.2cm}
{\cal R}^{2}_{13} + {\cal J}^{2}_\nu \; ,
\nonumber \\
{\cal S}^{2}_2 {\cal S}^{2}_3 \hspace{-0.2cm} & = & \hspace{-0.2cm}
{\cal R}^{2}_{23} + {\cal J}^{2}_\nu \; . \hspace{0.4cm}
\label{5}
\end{eqnarray}
On the other hand, taking the real part of each of the three equations in
Eq.~(\ref{3}) allows us to get at the following simple relations between
${\cal S}^{}_i$ and ${\cal R}^{}_{ij}$:
\begin{eqnarray}
{\cal S}^{2}_1 \hspace{-0.2cm} & = & \hspace{-0.2cm}
-{\cal R}^{}_{12} - {\cal R}^{}_{13} \; ,
\nonumber \\
{\cal S}^{2}_2 \hspace{-0.2cm} & = & \hspace{-0.2cm}
-{\cal R}^{}_{23} - {\cal R}^{}_{12} \; ,
\nonumber \\
{\cal S}^{2}_3 \hspace{-0.2cm} & = & \hspace{-0.2cm}
-{\cal R}^{}_{13} - {\cal R}^{}_{23} \; , \hspace{0.4cm}
\label{6}
\end{eqnarray}
which can also be seen from Fig.~\ref{fig1} or Fig.~\ref{fig2} in
a direct way. Substituting Eq.~(\ref{6}) into Eq.~(\ref{5}), we immediately
achieve the novel result
\begin{eqnarray}
{\cal J}^{2}_\nu = {\cal R}^{}_{12} {\cal R}^{}_{13} +
{\cal R}^{}_{12} {\cal R}^{}_{23} + {\cal R}^{}_{13}
{\cal R}^{}_{23} \; .
\label{7}
\end{eqnarray}
In other words, the CP-violating quantity ${\cal J}^{}_\nu$ and the
CP-conserving quantities ${\cal R}^{}_{ij}$ satisfy a Pythagoras-like
theorem in the three-flavor space where the observable effects of CP
violation in neutrino oscillations must require the participation of
all the three neutrino flavors. Note that
${\cal J}^{}_\nu$, ${\cal R}^{}_{12}$, ${\cal R}^{}_{13}$ and
${\cal R}^{}_{23}$ can all be extracted from a {\it single} neutrino
oscillation channel $\nu^{}_\mu \to \nu^{}_e$ or its CP-conjugated
process $\overline{\nu}^{}_\mu \to \overline{\nu}^{}_e$, at least
in principle. This nontrivial observation makes it possible to cross
check the magnitude of ${\cal J}^{}_\nu$ that will be directly
measured in the upcoming long-baseline experiment by calculating it
with the help of the Pythagoras-like relation in Eq.~(\ref{7}).

So far we have assumed the $3\times 3$ PMNS matrix $U$ to be exactly
unitary. This is certainly a very reasonable approximation, as
a careful analysis of the currently available electroweak precision
measurements and neutrino oscillation data has well constrained
a possible departure of $U U^\dagger$ or $U^\dagger U$ from the
identity matrix $I$ --- such a departure is expected
to be at most of ${\cal O}\left(\lesssim 10^{-2}\right)$~\cite{Antusch:2006vwa,
Antusch:2009gn,Blennow:2016jkn,Hu:2020oba,Wang:2021rsi,Blennow:2023mqx}.
Let us consider the possibility that the orthogonality relation of
$U$ in Eq.~(\ref{2}) is slightly violated in the type-I seesaw mechanism,
\begin{eqnarray}
U^{}_{e 1} U^*_{\mu 1} + U^{}_{e 2} U^*_{\mu 2} + U^{}_{e 3} U^*_{\mu 3}
= \epsilon \; ,
\label{8}
\end{eqnarray}
where $\epsilon$ denotes a small complex quantity characterizing the
active-sterile flavor mixing effect associated with the electron
and muon flavor indices. In this case one may reformulate the
the probabilities of $\nu^{}_\mu \to \nu^{}_e$ and $\overline{\nu}^{}_\mu
\to \overline{\nu}^{}_e$ oscillations and arrive at~\cite{Xing:2007zj,Li:2015oal}
\begin{eqnarray}
P(\nu^{}_\mu \to \nu^{}_e) \hspace{-0.2cm} & = & \hspace{-0.2cm}
\frac{\displaystyle \left | \epsilon \right |^{2}
- 4 \sum_{i<j} \left({\cal R}^{}_{ij} \sin^2 \frac{\Delta^{}_{ji} L}{4 E} \right)
+ 2 \sum_{i<j} \left( {\cal J}^{}_{ij} \sin \frac{\Delta^{}_{ji} L}{2 E} \right ) }
{\displaystyle \sum_{i} |U^{}_{e i}|^{2} \sum_{i} |U^{}_{\mu i}|^{2}} \; ,
\nonumber \\
P(\overline{\nu}^{}_\mu \to \overline{\nu}^{}_e) \hspace{-0.2cm} & = &
\hspace{-0.2cm}
\frac{\displaystyle \left | \epsilon \right |^{2}
- 4 \sum_{i<j} \left({\cal R}^{}_{ij} \sin^2 \frac{\Delta^{}_{ji} L}{4 E} \right)
- 2 \sum_{i<j} \left( {\cal J}^{}_{ij} \sin \frac{\Delta^{}_{ji} L}{2 E} \right ) }
{\displaystyle \sum_{i} |U^{}_{e i}|^{2} \sum_{i} |U^{}_{\mu i}|^{2}} \; ,
\hspace{0.4cm}
\label{9}
\end{eqnarray}
where ${\cal R}^{}_{ij} \equiv {\rm Re} (U^{}_{\mu i} U^{}_{e j} U^*_{\mu j}
U^*_{e i})$ and ${\cal J}^{}_{ij} \equiv {\rm Im} (U^{}_{\mu i} U^{}_{e j}
U^*_{\mu j} U^*_{e i})$ (for $i, j = 1, 2, 3$ and $i < j$).
We see that $|\epsilon|^2 \neq 0$ actually measures the zero-distance effect of
$\nu^{}_\mu \to \nu^{}_e$ and $\overline{\nu}^{}_\mu \to \overline{\nu}^{}_e$
flavor conversions at $L/E = 0$. It is easy to
show that ${\cal J}^{}_{12}$, ${\cal J}^{}_{13}$ and ${\cal J}^{}_{23}$
are correlated with one another as follows:
\begin{eqnarray}
{\cal J}^{}_{12} + {\cal J}^{}_{13} \hspace{-0.2cm} & = & \hspace{-0.2cm}
+ {\rm Im} ( \epsilon \; U^{*}_{e 1} U^{}_{\mu 1} ) \; ,
\nonumber \\
{\cal J}^{}_{12} - {\cal J}^{}_{23} \hspace{-0.2cm} & = & \hspace{-0.2cm}
- {\rm Im} ( \epsilon \; U^{*}_{e 2} U^{}_{\mu 2} ) \; ,
\nonumber \\
{\cal J}^{}_{13} + {\cal J}^{}_{23} \hspace{-0.2cm} & = & \hspace{-0.2cm}
- {\rm Im} ( \epsilon \; U^{*}_{e 3} U^{}_{\mu 3} ) \; ; \hspace{0.4cm}
\label{10}
\end{eqnarray}
while ${\cal R}^{}_{12}$, ${\cal R}^{}_{13}$ and ${\cal R}^{}_{23}$
are correlated with one another through the relations
\begin{eqnarray}
\left|U^{}_{e 1} U^*_{\mu 1}\right|^2 + {\cal R}^{}_{12} + {\cal R}^{}_{13}
\hspace{-0.2cm} & = & \hspace{-0.2cm}
{\rm Re} ( \epsilon \; U^{*}_{e 1} U^{}_{\mu 1} ) \; ,
\nonumber \\
\left|U^{}_{e 2} U^*_{\mu 2}\right|^2 + {\cal R}^{}_{12} + {\cal R}^{}_{23}
\hspace{-0.2cm} & = & \hspace{-0.2cm}
{\rm Re} ( \epsilon \; U^{*}_{e 2} U^{}_{\mu 2} ) \; ,
\nonumber \\
\left|U^{}_{e 3} U^*_{\mu 3}\right|^2 + {\cal R}^{}_{13} + {\cal R}^{}_{23}
\hspace{-0.2cm} & = & \hspace{-0.2cm}
{\rm Re} ( \epsilon \; U^{*}_{e 3} U^{}_{\mu 3} ) \; . \hspace{0.4cm}
\label{11}
\end{eqnarray}
So the exact unitarity of $U$ (i.e., $\epsilon = 0$) leads us to
${\cal J}^{}_{12} = -{\cal J}^{}_{13} = {\cal J}^{}_{23} \equiv -{\cal J}^{}_\nu$,
and then Eq.~(\ref{9}) will automatically be reduced to Eq.~(\ref{1}).
Furthermore, we find
\begin{eqnarray}
{\cal J}^{2}_{12} \hspace{-0.2cm} & = & \hspace{-0.2cm}
\left|U^{}_{e 1} U^*_{\mu 1}\right|^2 \left|U^{}_{e 2} U^*_{\mu 2}\right|^2
- {\cal R}^{2}_{12}
\nonumber \\
\hspace{-0.2cm} & = & \hspace{-0.2cm}
{\cal R}^{}_{12} {\cal R}^{}_{13}
+ {\cal R}^{}_{12} {\cal R}^{}_{23} + {\cal R}^{}_{13} {\cal R}^{}_{23}
- \Big[ {\rm Re} ( \epsilon \; U^{*}_{e 1} U^{}_{\mu 1} )
\left ( {\cal R}^{}_{12} + {\cal R}^{}_{23} \right )
\nonumber \\
&& \hspace{-0.2cm} + \hspace{0.06cm} {\rm Re} ( \epsilon \; U^{*}_{e 2} U^{}_{\mu 2} )
\left ( {\cal R}^{}_{12} + {\cal R}^{}_{13} \right )
- {\rm Re} ( \epsilon \; U^{*}_{e 1} U^{}_{\mu 1} )
{\rm Re} ( \epsilon \; U^{*}_{e 2} U^{}_{\mu 2} ) \Big] \; , \hspace{0.4cm}
\nonumber\\
{\cal J}^{2}_{13} \hspace{-0.2cm} & = & \hspace{-0.2cm}
\left|U^{}_{e 1} U^*_{\mu 1}\right|^2 \left|U^{}_{e 3} U^*_{\mu 3}\right|^2
- {\cal R}^{2}_{13}
\nonumber \\
\hspace{-0.2cm} & = & \hspace{-0.2cm}
{\cal R}^{}_{12} {\cal R}^{}_{13}
+ {\cal R}^{}_{12} {\cal R}^{}_{23} + {\cal R}^{}_{13} {\cal R}^{}_{23}
- \Big[ {\rm Re} ( \epsilon \; U^{*}_{e 1} U^{}_{\mu 1} )
\left ( {\cal R}^{}_{13} + {\cal R}^{}_{23} \right )
\nonumber \\
&& \hspace{-0.2cm} + \hspace{0.06cm} {\rm Re} ( \epsilon \; U^{*}_{e 3} U^{}_{\mu 3} )
\left ( {\cal R}^{}_{12} + {\cal R}^{}_{13} \right )
- {\rm Re} ( \epsilon \; U^{*}_{e 1} U^{}_{\mu 1} )
{\rm Re} ( \epsilon \; U^{*}_{e 3} U^{}_{\mu 3} ) \Big] \; ,
\nonumber\\
{\cal J}^{2}_{23} \hspace{-0.2cm} & = & \hspace{-0.2cm}
\left|U^{}_{e 2} U^*_{\mu 2}\right|^2 \left|U^{}_{e 3} U^*_{\mu 3}\right|^2
- {\cal R}^{2}_{23}
\nonumber \\
\hspace{-0.2cm} & = & \hspace{-0.2cm}
{\cal R}^{}_{12} {\cal R}^{}_{13}
+ {\cal R}^{}_{12} {\cal R}^{}_{23} + {\cal R}^{}_{13} {\cal R}^{}_{23}
- \Big[ {\rm Re} ( \epsilon \; U^{*}_{e 2} U^{}_{\mu 2} )
\left ( {\cal R}^{}_{13} + {\cal R}^{}_{23} \right )
\nonumber \\
&& \hspace{-0.2cm} + \hspace{0.06cm} {\rm Re} ( \epsilon \; U^{*}_{e 3} U^{}_{\mu 3} )
\left ( {\cal R}^{}_{12} + {\cal R}^{}_{23} \right )
- {\rm Re} ( \epsilon \; U^{*}_{e 2} U^{}_{\mu 2} )
{\rm Re} ( \epsilon \; U^{*}_{e 3} U^{}_{\mu 3} ) \Big] \; ,
\label{12}
\end{eqnarray}
from which the universality relation ${\cal J}^2_{12} = {\cal J}^2_{13} =
{\cal J}^2_{23} = {\cal J}^2_\nu$ and the Pythagoras-like relation
obtained in Eq.~(\ref{7}) can simply be reproduced from $\epsilon = 0$
in the unitarity limit of $U$. Such an example illustrates that an
experimental test of the validity of Eq.~(\ref{7}) may allow us to
diagnose possible non-unitarity of the $3\times 3$ PMNS matrix $U$.

\section{Terrestrial matter effects}

We proceed to discuss terrestrial matter effects on ${\cal J}^{}_\nu$ and
${\cal R}^{}_{ij}$ (for $ij = 12, 13, 23$) in the unitarity limit of $U$,
since such effects are inevitable in a realistic long-baseline experiment of
$\nu^{}_\mu \to \nu^{}_e$ and $\overline{\nu}^{}_\mu \to \overline{\nu}^{}_e$
oscillations. In this case the corresponding probabilities can be simply
expressed as
\begin{eqnarray}
\widetilde{P}(\nu^{}_\mu \to \nu^{}_e) \hspace{-0.2cm} & = & \hspace{-0.2cm}
- 4 \sum_{i<j} \left(\widetilde{\cal R}^{}_{ij} \sin^2
\frac{\widetilde{\Delta}^{}_{ji} L}{4 E} \right)
- 8 \widetilde{\cal J}^{}_\nu \prod_{i<j}
\sin \frac{\widetilde{\Delta}^{}_{ji} L}{4 E} \; ,
\nonumber \\
\widetilde{P}(\overline{\nu}^{}_\mu \to \overline{\nu}^{}_e)
\hspace{-0.2cm} & = & \hspace{-0.2cm}
- 4 \sum_{i<j} \left(\widetilde{\cal R}^{}_{ij} \sin^2
\frac{\widetilde{\Delta}^{}_{ji} L}{4 E} \right)
+ 8 \widetilde{\cal J}^{}_\nu \prod_{i<j}
\sin \frac{\widetilde{\Delta}^{}_{ji} L}{4 E} \; , \hspace{0.4cm}
\label{13}
\end{eqnarray}
where $\widetilde{\cal R}^{}_{ij} = {\rm Re}
(\widetilde{U}^{}_{\mu i} \widetilde{U}^{}_{e j}
\widetilde{U}^*_{\mu j} \widetilde{U}^*_{e i})$,
$\widetilde{\cal J}^{}_\nu = \left(-1\right)^{i+j} \hspace{0.05cm}
{\rm Im} (\widetilde{U}^{}_{\mu i}
\widetilde{U}^{}_{e j} \widetilde{U}^*_{\mu j} \widetilde{U}^*_{e i})$,
$\widetilde{\Delta}^{}_{ji} \equiv \widetilde{m}^2_j - \widetilde{m}^2_i$
and $\widetilde{U}$ are just the respective analogues of ${\cal R}^{}_{ij}$,
${\cal J}^{}_\nu$, $\Delta^{}_{ji}$ and $U$ in matter (for
$i, j = 1, 2, 3$ and $i < j$). These effective quantities depend on
the matter parameter $A = 2\sqrt{2} \hspace{0.05cm} G^{}_{\rm F} N^{}_e E$
characterizing the strength of elastic forward coherent scattering of $\nu^{}_e$
events with the electrons of a number density $N^{}_e$ in the earth via
the weak charged-current interactions~\cite{Wolfenstein:1977ue,Mikheyev:1985zog},
and $A$ must flip its sign for the probability of $\overline{\nu}^{}_\mu \to
\overline{\nu}^{}_e$ oscillations in matter.
The explicit expressions of $\widetilde{\Delta}^{}_{ji}$ and
$\Delta^\prime_{ii} \equiv \widetilde{m}^2_i - m^2_i$ have already
been given in Refs.~\cite{Barger:1980tf,Xing:2000gg,Xing:2003ez}.
After a lengthy but straightforward calculation, we find that
$\widetilde{\cal R}^{}_{ij}$ and ${\cal R}^{}_{ij}$ are related to each other through
\begin{eqnarray}
\widetilde{\cal R}^{}_{12} \hspace{-0.2cm} & = & \hspace{-0.2cm}
\frac{\Delta^2_{21} \left(\Delta^{}_{31} - \Delta^\prime_{11}\right)
\left(\Delta^{}_{32} - \Delta^\prime_{22}\right)}
{\widetilde{\Delta}^2_{21} \widetilde{\Delta}^{}_{31} \widetilde{\Delta}^{}_{32}}
{\cal R}^{}_{12} -
\frac{\Delta^2_{31} \Delta^\prime_{22}
\left(\Delta^{}_{21} - \Delta^\prime_{11}\right)}
{\widetilde{\Delta}^2_{21} \widetilde{\Delta}^{}_{31} \widetilde{\Delta}^{}_{32}}
{\cal R}^{}_{13} \hspace{0.4cm}
\nonumber \\
\hspace{-0.2cm} & & \hspace{-0.2cm}
+ \frac{\Delta^2_{32} \Delta^\prime_{11}
\left(\Delta^{}_{21} + \Delta^\prime_{22}\right)}
{\widetilde{\Delta}^2_{21} \widetilde{\Delta}^{}_{31} \widetilde{\Delta}^{}_{32}}
{\cal R}^{}_{23} \; ,
\nonumber \\
\widetilde{\cal R}^{}_{13} \hspace{-0.2cm} & = & \hspace{-0.2cm}
\frac{\Delta^2_{31} \left(\Delta^{}_{21} - \Delta^\prime_{11}\right)
\left(\Delta^{}_{32} + \Delta^\prime_{33}\right)}
{\widetilde{\Delta}^{2}_{31} \widetilde{\Delta}^{}_{21} \widetilde{\Delta}^{}_{32}}
{\cal R}^{}_{13} +
\frac{\Delta^2_{21} \Delta^\prime_{33}
\left(\Delta^{}_{31} - \Delta^\prime_{11}\right)}
{\widetilde{\Delta}^{2}_{31} \widetilde{\Delta}^{}_{21} \widetilde{\Delta}^{}_{32}}
{\cal R}^{}_{12} \hspace{0.4cm}
\nonumber \\
\hspace{-0.2cm} & & \hspace{-0.2cm}
- \frac{\Delta^2_{32} \Delta^\prime_{11}
\left(\Delta^{}_{31} + \Delta^\prime_{33}\right)}
{\widetilde{\Delta}^{2}_{31} \widetilde{\Delta}^{}_{21} \widetilde{\Delta}^{}_{32}}
{\cal R}^{}_{23} \; ,
\nonumber \\
\widetilde{\cal R}^{}_{23} \hspace{-0.2cm} & = & \hspace{-0.2cm}
\frac{\Delta^2_{32} \left(\Delta^{}_{21} + \Delta^\prime_{22}\right)
\left(\Delta^{}_{31} + \Delta^\prime_{33}\right)}
{\widetilde{\Delta}^{2}_{32} \widetilde{\Delta}^{}_{21} \widetilde{\Delta}^{}_{31}}
{\cal R}^{}_{23} -
\frac{\Delta^2_{21} \Delta^\prime_{33}
\left(\Delta^{}_{32} - \Delta^\prime_{22}\right)}
{\widetilde{\Delta}^{2}_{32} \widetilde{\Delta}^{}_{21} \widetilde{\Delta}^{}_{31}}
{\cal R}^{}_{12} \hspace{0.4cm}
\nonumber \\
\hspace{-0.2cm} & & \hspace{-0.2cm}
+ \frac{\Delta^2_{31} \Delta^\prime_{22}
\left(\Delta^{}_{32} + \Delta^\prime_{33}\right)}
{\widetilde{\Delta}^{2}_{32} \widetilde{\Delta}^{}_{21} \widetilde{\Delta}^{}_{31}}
{\cal R}^{}_{13} \; .
\label{14}
\end{eqnarray}
On the other hand, $\widetilde{\cal J}^{}_\nu$ and ${\cal J}^{}_\nu$
satisfy the well-known Naumov relation~\cite{Naumov:1991ju}
\begin{eqnarray}
\widetilde{\cal J}^{}_\nu = \frac{\Delta^{}_{21} \Delta^{}_{31}
\Delta^{}_{32}}{\widetilde{\Delta}^{}_{21} \widetilde{\Delta}^{}_{31}
\widetilde{\Delta}^{}_{32}} {\cal J}^{}_\nu \; .
\label{15}
\end{eqnarray}
The unitarity of $\widetilde{U}$ assures that a Pythagoras-like relation similar
to Eq.~(\ref{7}) exists between $\widetilde{\cal R}^{}_{ij}$
and $\widetilde{\cal J}^{}_\nu$ as follows:
\begin{eqnarray}
\widetilde{\cal J}^{2}_\nu = \widetilde{\cal R}^{}_{12}
\widetilde{\cal R}^{}_{13} +
\widetilde{\cal R}^{}_{12} \widetilde{\cal R}^{}_{23} +
\widetilde{\cal R}^{}_{13} \widetilde{\cal R}^{}_{23} \; .
\label{16}
\end{eqnarray}
This relation provides a cross-check of the matter-corrected CP-violating effect
that can be directly measured in a realistic long-baseline $\nu^{}_\mu \to \nu^{}_e$ and
$\overline{\nu}^{}_\mu \to \overline{\nu}^{}_e$ oscillations by calculating it
from the associated parameters $\widetilde{\cal R}^{}_{ij}$ via Eq.~(\ref{16}).

To illustrate how significant (or insignificant) the terrestrial matter effects on
${\cal R}^{}_{ij}$ and ${\cal J}^{}_\nu$ for the T2K~\cite{T2K:2023smv},
Hyper-K~\cite{Hyper-KamiokandeProto-:2015xww},
T2HKK~\cite{Hyper-Kamiokande:2016srs} and DUNE~\cite{DUNE:2020jqi} experiments,
we simply assume a constant earth matter density $\rho = 2.8 ~{\rm g}/{\rm cm}^3$
and an electron number fraction $Y^{}_e = 0.5$ to estimate the size of $A$ as a
function of $E$~\cite{Mocioiu:2000st}:
$A \simeq 1.52 \times 10^{-4} ~{\rm eV}^2 ~Y^{}_e \left(\rho/{\rm g}/{\rm cm}^3
\right) \left(E/{\rm GeV}\right) \simeq 2.13 \times 10^{-4} ~{\rm eV}^2
\left(E/{\rm GeV}\right)$. The best-fit values of $\Delta^{}_{21}$, $\Delta^{}_{31}$,
$\theta^{}_{12}$, $\theta^{}_{13}$, $\theta^{}_{23}$ and $\delta^{}_\nu$ given
in Refs.~\cite{Gonzalez-Garcia:2021dve,Nufit} are used in our
calculations as the typical
inputs. Our numerical results are summarized in Fig.~\ref{3} for the
normal neutrino mass ordering and in Fig.~\ref{4} for the
inverted neutrino mass ordering. Some brief discussions are in order.
\begin{figure}[t]
\begin{center}
\vspace{-0.2cm}
\includegraphics[width=14.2cm]{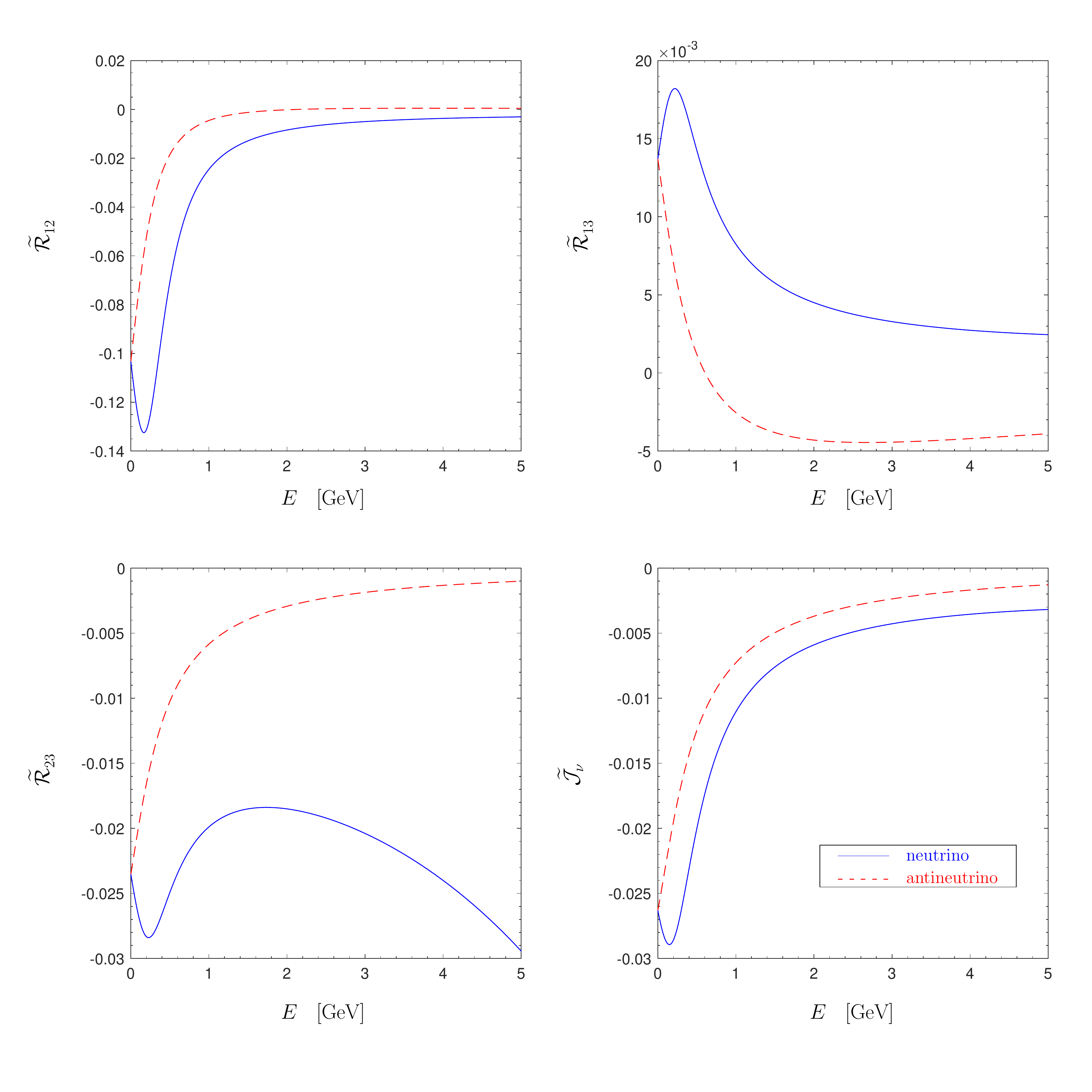}
\vspace{-1.cm}
\caption{The changes of $\widetilde{\cal R}^{}_{ij}$ (for $ij = 12, 13, 23$) and
$\widetilde{\cal J}^{}_\nu$ as functions of the neutrino (or antineutrino) beam energy $E$,
where a constant earth matter density profile with $\rho = 2.8 ~{\rm g} / {\rm cm}^{3}_{}$
and $Y^{}_{e} = 0.5$ have been assumed and the best-fit values of the relevant neutrino
oscillation parameters in the {\bf normal} mass ordering
case~\cite{Gonzalez-Garcia:2021dve,Nufit} have been input.}
\label{fig3}
\end{center}
\end{figure}
\begin{figure}[t]
\begin{center}
\vspace{-0.2cm}
\includegraphics[width=14.2cm]{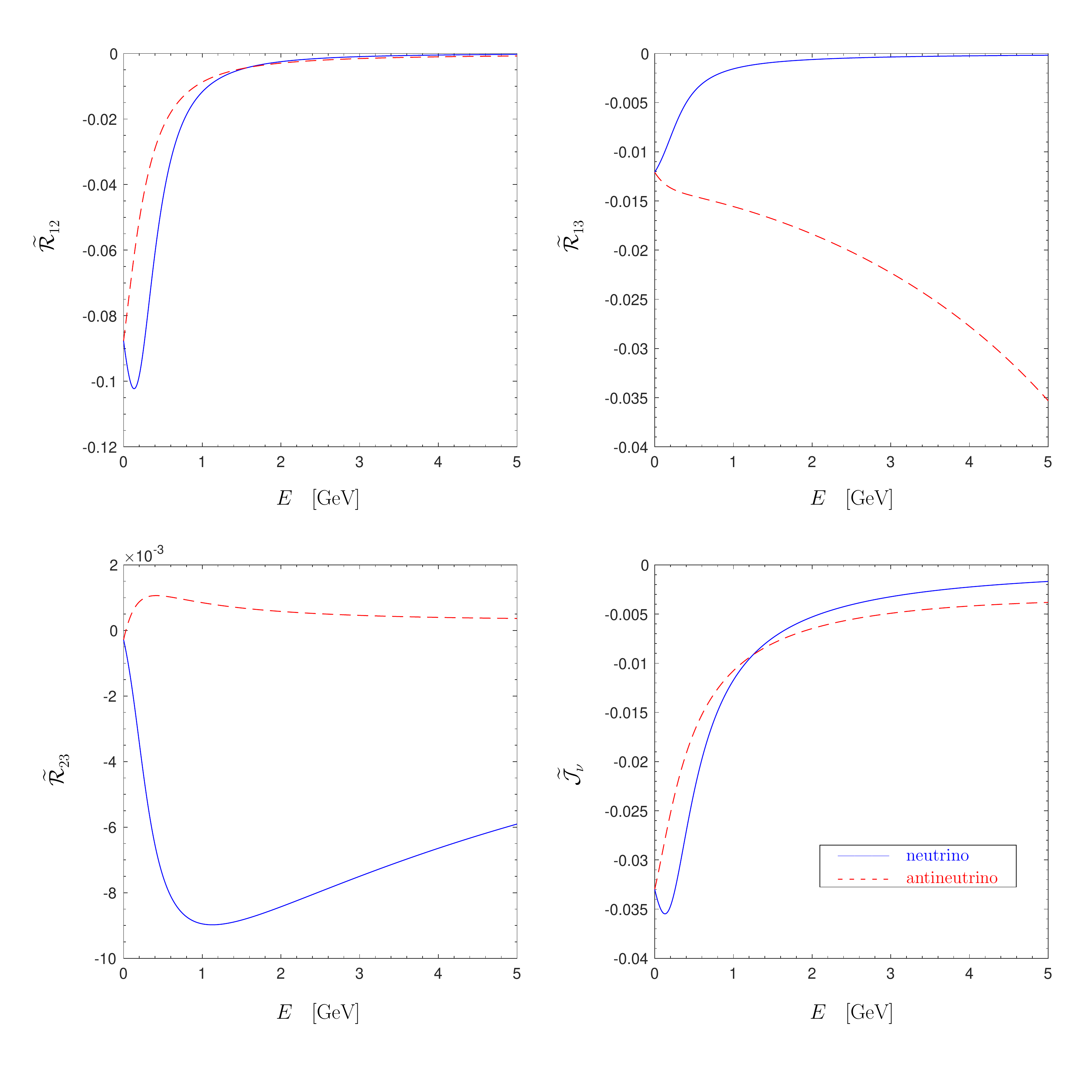}
\vspace{-1.cm}
\caption{The changes of $\widetilde{\cal R}^{}_{ij}$ (for $ij = 12, 13, 23$) and
$\widetilde{\cal J}^{}_\nu$ as functions of the neutrino (or antineutrino) beam energy $E$,
where a constant earth matter density profile with $\rho = 2.8 ~{\rm g} / {\rm cm}^{3}_{}$
and $Y^{}_{e} = 0.5$ have been assumed and the best-fit values of the relevant neutrino
oscillation parameters in the {\bf inverted} mass ordering
case~\cite{Gonzalez-Garcia:2021dve,Nufit} have been input.}
\label{fig3}
\end{center}
\end{figure}

(1) A general observation is that both ${\cal R}^{}_{ij}$ and ${\cal J}^{}_\nu$
are quite sensitive to the matter-induced corrections, and the latter may even flip the
signs of the former, depending on whether the neutrino or antineutrino beam is
under consideration and whether the neutrino mass ordering is normal or inverted. So
it makes sense to take into account the terrestrial matter effects when applying the
Pythagoras-like relation in Eq.~(\ref{7}) to the description of CP violation in
$\nu^{}_\mu \to \nu^{}_e$ and $\overline{\nu}^{}_\mu \to \overline{\nu}^{}_e$ oscillations
for a realistic long-baseline experiment.

(2) As compared with ${\cal J}^{}_\nu$, the magnitude of $\widetilde{\cal J}^{}_\nu$
undergoes a small enhancement for the {\it neutrino} beam with $E < 1 ~{\rm GeV}$, no
matter whether the neutrino mass spectrum is normal or inverted.
Such an interesting behavior has been understood in a reliable analytical
approximation as done in Ref.~\cite{Xing:2016ymg}. The resonant enhancement
$\widetilde{\cal J}^{}_\nu / {\cal J}^{}_\nu \simeq 1/\sin 2\theta^{}_{12}
\simeq 1.1$ is found to appear at
$E \simeq \Delta^{}_{21} \cos 2\theta^{}_{12}/\left(2 \sqrt{2} \hspace{0.07cm}
G^{}_{\rm F} N^{}_e\right) \simeq 0.14 ~{\rm GeV}$ to the leading-order degree
of accuracy, which is essentially consistent with the numerical result shown in
Fig.~\ref{3} or Fig.~\ref{4}. As for a low-energy antineutrino beam, there is no
matter-induced resonance for $\widetilde{\cal J}^{}_\nu$.

(3) The behavior of $\widetilde{\cal R}^{}_{12}$ changing with the neutrino (or antineutrino)
beam energy $E$ is rather similar to that of $\widetilde{\cal J}^{}_\nu$, especially
in the $E \lesssim 1~{\rm GeV}$ region. This observation implies that both in vacuum
and in low-density matter with $E \lesssim 1~{\rm GeV}$, ${\cal R}^{}_{12}$ and its
counterpart $\widetilde{\cal R}^{}_{12}$ play the dominant role in the corresponding
Pythagoras-like relations given respectively by Eq.~(\ref{7}) and Eq.~(\ref{16}).
In comparison, $\widetilde{\cal R}^{}_{13}$ and $\widetilde{\cal R}^{}_{23}$ may
also undergo a local resonance when $E$ increases, and their explicit running
behaviors with $E$ can similarly be understood by doing an analogous analytical
approximation as in Ref.~\cite{Xing:2016ymg}.

In the presence of terrestrial matter effects, one may similarly consider the
possibility of slight unitarity violation of the $3\times 3$ PMNS matrix $U$ or
its matter-corrected counterpart $\widetilde{U}$, and then follow the
procedures outlined in Eqs.~(\ref{8})---(\ref{12}) to discuss how the effective
Pythagoras-like relation obtained in Eq.~(\ref{16}) can be modified. Here we
do not repeat such calculations.

\section{Summary}

The most important goal of the next-generation long-baseline neutrino oscillation
experiments, which all aim to precisely measure the {\it appearance}
oscillation channels $\nu^{}_\mu \to \nu^{}_e$ and $\overline{\nu}^{}_\mu \to
\overline{\nu}^{}_e$, is to convincingly establish leptonic CP violation that
is elegantly characterized by the Jarlskog invariant ${\cal J}^{}_\nu$ in vacuum
or its counterpart $\widetilde{\cal J}^{}_\nu$ in matter. In this work we have found a
Pythagoras-like theorem for the expression of ${\cal J}^{}_\nu$ in terms of the
CP-conserving quantities ${\cal R}^{}_{12}$, ${\cal R}^{}_{13}$ and ${\cal R}^{}_{23}$,
which also appear in the probabilities of $\nu^{}_\mu \to \nu^{}_e$ and
$\overline{\nu}^{}_\mu \to \overline{\nu}^{}_e$ oscillations.
Indirect non-unitarity effects on ${\cal J}^{}_\nu$ and ${\cal R}^{}_{ij}$ have
also been discussed. The same
Pythagoras-like relation holds for the matter-corrected quantities
$\widetilde{\cal J}^{}_\nu$ and $\widetilde{\cal R}^{}_{ij}$ (for
$ij = 12, 13, 23$), as a natural consequence of the unitarity of the effective
PMNS matrix $\widetilde{U}$. Namely, the magnitude of ${\cal J}^{}_\nu$
(or $\widetilde{\cal J}^{}_\nu$) can be simply calculated from the precision
measurements of ${\cal R}^{}_{ij}$ (or $\widetilde{\cal R}^{}_{ij}$) with the
help of the Pythagoras-like relation. This finding may hopefully provide
a novel cross-check of the result of ${\cal J}^{}_\nu$ that will be directly
measured in the near future.

We expect that such a Pythagoras-like theorem can be simply extended to study
leptonic CP violation and test unitarity of the PMNS matrix in some
other channels of neutrino oscillations, such as
$\nu^{}_\mu \to \nu^{}_\tau$ and $\overline{\nu}^{}_\mu \to
\overline{\nu}^{}_\tau$ or $\nu^{}_e \to \nu^{}_\tau$ and
$\overline{\nu}^{}_e \to \overline{\nu}^{}_\tau$ oscillations. It is also possible
to apply a similar rephasing-invariant language to the
description of CP violation in the quark sector. Note, however, that the observable
effects of CP violation in various quark decay modes are usually characterized by
the ratios ${\cal J}^{}_q /{\rm Re}(V^{}_{\alpha i} V^{}_{\beta j} V^*_{\alpha j}
V^*_{\beta i})$, where ${\cal J}^{}_q$ denotes the corresponding Jarlskog invariant,
$\alpha$ and $\beta$ (or $i$ and $j$) are the up-type (or down-type)
quark flavor indices which run over $u$, $c$ and $t$ (or $d$, $s$ and $b$).
A further study along this line of thought is beyond the scope of this paper
and will be presented elsewhere.

\vspace{0.5cm}

This work is supported in part by the National Natural Science Foundation of
China under grant No. 11775183 (S.L.) and grant Nos. 12075254 and
11835013 (Z.Z.X.).


\begin{thebibliography}{99}
\bibitem{ParticleDataGroup:2022pth}
R.~L.~Workman \textit{et al.} [Particle Data Group],
``Review of Particle Physics,''
PTEP \textbf{2022} (2022), 083C01.

\bibitem{Pontecorvo:1957cp}
B.~Pontecorvo,
``Mesonium and anti-mesonium,''
Sov.\ Phys.\ JETP {\bf 6} (1957) 429
[Zh.\ Eksp.\ Teor.\ Fiz.\  {\bf 33} (1957) 549].

\bibitem{Maki:1962mu}
Z.~Maki, M.~Nakagawa and S.~Sakata,
``Remarks on the unified model of elementary particles,''
Prog.\ Theor.\ Phys.\  {\bf 28} (1962) 870.

\bibitem{Pontecorvo:1967fh}
B.~Pontecorvo,
``Neutrino Experiments and the Problem of Conservation of Leptonic Charge,''
Sov.\ Phys.\ JETP {\bf 26} (1968) 984
[Zh.\ Eksp.\ Teor.\ Fiz.\  {\bf 53} (1967) 1717].

\bibitem{Antusch:2006vwa}
S.~Antusch, C.~Biggio, E.~Fernandez-Martinez, M.~B.~Gavela and J.~Lopez-Pavon,
``Unitarity of the Leptonic Mixing Matrix,''
JHEP \textbf{10} (2006), 084
[arXiv:hep-ph/0607020 [hep-ph]].

\bibitem{Antusch:2009gn}
S.~Antusch, S.~Blanchet, M.~Blennow and E.~Fernandez-Martinez,
``Non-unitary Leptonic Mixing and Leptogenesis,''
JHEP \textbf{01} (2010), 017
[arXiv:0910.5957 [hep-ph]].

\bibitem{Blennow:2016jkn}
M.~Blennow, P.~Coloma, E.~Fernandez-Martinez, J.~Hernandez-Garcia and J.~Lopez-Pavon,
``Non-Unitarity, sterile neutrinos, and Non-Standard neutrino Interactions,''
JHEP \textbf{04} (2017), 153
[arXiv:1609.08637 [hep-ph]].

\bibitem{Hu:2020oba}
Z.~Hu, J.~Ling, J.~Tang and T.~Wang,
``Global oscillation data analysis on the $3\nu$ mixing without unitarity,''
JHEP \textbf{01} (2021), 124
[arXiv:2008.09730 [hep-ph]].

\bibitem{Wang:2021rsi}
Y.~Wang and S.~Zhou,
``Non-unitary leptonic flavor mixing and CP violation in neutrino-antineutrino oscillations,''
Phys. Lett. B \textbf{824} (2022), 136797
[arXiv:2109.13622 [hep-ph]].

\bibitem{Blennow:2023mqx}
M.~Blennow, E.~Fern\'andez-Mart\'\i{}nez, J.~Hern\'andez-Garc\'\i{}a, J.~L\'opez-Pav\'on, X.~Marcano and D.~Naredo-Tuero,
``Bounds on lepton non-unitarity and heavy neutrino mixing,''
[arXiv:2306.01040 [hep-ph]].

\bibitem{Konetschny:1977bn}
W.~Konetschny and W.~Kummer,
``Nonconservation of Total Lepton Number with Scalar Bosons,''
Phys. Lett. B \textbf{70} (1977), 433-435.

\bibitem{Magg:1980ut}
M.~Magg and C.~Wetterich,
``Neutrino Mass Problem and Gauge Hierarchy,''
Phys. Lett. B \textbf{94} (1980), 61-64.

\bibitem{Schechter:1980gr}
J.~Schechter and J.~W.~F.~Valle,
``Neutrino Masses in SU(2) x U(1) Theories,''
Phys. Rev. D \textbf{22} (1980), 2227.

\bibitem{Cheng:1980qt}
T.~P.~Cheng and L.~F.~Li,
``Neutrino Masses, Mixings and Oscillations in SU(2) x U(1) Models of Electroweak Interactions,''
Phys. Rev. D \textbf{22} (1980), 2860.

\bibitem{Lazarides:1980nt}
G.~Lazarides, Q.~Shafi and C.~Wetterich,
``Proton Lifetime and Fermion Masses in an SO(10) Model,''
Nucl. Phys. B \textbf{181} (1981), 287-300.

\bibitem{Mohapatra:1980yp}
R.~N.~Mohapatra and G.~Senjanovic,
``Neutrino Masses and Mixings in Gauge Models with Spontaneous Parity Violation,''
Phys. Rev. D \textbf{23} (1981), 165.

\bibitem{Minkowski:1977sc}
P.~Minkowski,
``$\mu \to e\gamma$ at a rate of one out of $10^{9}$ muon decays?,''
Phys.\ Lett.\  {\bf 67B} (1977) 421.

\bibitem{Yanagida:1979as}
T.~Yanagida,
``Horizontal gauge symmetry and masses of neutrinos,''
Conf.\ Proc.\ C {\bf 7902131} (1979) 95.

\bibitem{GellMann:1980vs}
M.~Gell-Mann, P.~Ramond and R.~Slansky,
``Complex spinors and unified theories,''
Conf.\ Proc.\ C {\bf 790927} (1979) 315
[arXiv:1306.4669 [hep-th]].

\bibitem{Glashow:1979nm}
S.~L.~Glashow,
``The future of elementary particle physics,''
NATO Sci.\ Ser.\ B {\bf 61} (1980) 687.

\bibitem{Mohapatra:1979ia}
R.~N.~Mohapatra and G.~Senjanovic,
``Neutrino mass and spontaneous parity nonconservation,''
Phys.\ Rev.\ Lett.\  {\bf 44} (1980) 912.

\bibitem{Foot:1988aq}
R.~Foot, H.~Lew, X.~G.~He and G.~C.~Joshi,
``Seesaw Neutrino Masses Induced by a Triplet of Leptons,''
Z. Phys. C \textbf{44} (1989), 441.

\bibitem{T2K:2019bcf}
K.~Abe \textit{et al.} [T2K],
``Constraint on the matter\textendash{}antimatter symmetry-violating phase in neutrino oscillations,''
Nature \textbf{580} (2020) no.7803, 339-344
[erratum: Nature \textbf{583} (2020) no.7814, E16]
[arXiv:1910.03887 [hep-ex]].

\bibitem{T2K:2023smv}
K.~Abe \textit{et al.} [T2K],
``Measurements of neutrino oscillation parameters from the T2K experiment using $3.6\times10^{21}$ protons on target,''
[arXiv:2303.03222 [hep-ex]].


\bibitem{Jarlskog:1985ht}
C.~Jarlskog,
``Commutator of the Quark Mass Matrices in the Standard Electroweak Model and a Measure of Maximal $CP$~Nonconservation,''
Phys. Rev. Lett. \textbf{55} (1985), 1039.

\bibitem{Wu:1985ea}
D.~D.~Wu,
``The Rephasing Invariants and CP,''
Phys. Rev. D \textbf{33} (1986), 860.

\bibitem{Cheng:1986in}
H.~Y.~Cheng,
``{Kobayashi-Maskawa} Type of Hard {CP} Violation Model With Three Generation Majorana Neutrinos,''
Phys. Rev. D \textbf{34} (1986), 2794.

\bibitem{Hyper-KamiokandeProto-:2015xww}
K.~Abe \textit{et al.} [Hyper-Kamiokande Proto-],
``Physics potential of a long-baseline neutrino oscillation experiment using a J-PARC neutrino beam and Hyper-Kamiokande,''
PTEP \textbf{2015} (2015), 053C02
[arXiv:1502.05199 [hep-ex]].

\bibitem{Hyper-Kamiokande:2016srs}
K.~Abe \textit{et al.} [Hyper-Kamiokande],
``Physics potentials with the second Hyper-Kamiokande detector in Korea,''
PTEP \textbf{2018} (2018) no.6, 063C01
[arXiv:1611.06118 [hep-ex]].

\bibitem{DUNE:2020jqi}
B.~Abi \textit{et al.} [DUNE],
``Long-baseline neutrino oscillation physics potential of the DUNE experiment,''
Eur. Phys. J. C \textbf{80} (2020) no.10, 978
[arXiv:2006.16043 [hep-ex]].

\bibitem{Fritzsch:1999ee}
H.~Fritzsch and Z.~Z.~Xing,
``Mass and flavor mixing schemes of quarks and leptons,''
Prog. Part. Nucl. Phys. \textbf{45} (2000), 1-81
[arXiv:hep-ph/9912358 [hep-ph]].


\bibitem{Gonzalez-Garcia:2021dve}
M.~C.~Gonzalez-Garcia, M.~Maltoni and T.~Schwetz,
``NuFIT: Three-Flavour Global Analyses of Neutrino Oscillation Experiments,''
Universe \textbf{7} (2021) no.12, 459
[arXiv:2111.03086 [hep-ph]].


\bibitem{Nufit}
NuFit webpage, http://www.nu-fit.org.

\bibitem{Xing:2012zv}
Z.~Z.~Xing,
``Model-independent access to the structure of quark flavor mixing,''
Phys. Rev. D \textbf{86} (2012), 113006
[arXiv:1211.3890 [hep-ph]].


\bibitem{Xing:2007zj}
Z.~z.~Xing,
``Correlation between the Charged Current Interactions of Light and Heavy Majorana Neutrinos,''
Phys. Lett. B \textbf{660} (2008), 515-521
[arXiv:0709.2220 [hep-ph]].

\bibitem{Li:2015oal}
Y.~F.~Li and S.~Luo,
``Neutrino Oscillation Probabilities in Matter with Direct and Indirect Unitarity Violation in the Lepton Mixing Matrix,''
Phys. Rev. D \textbf{93} (2016) no.3, 033008
[arXiv:1508.00052 [hep-ph]].

\bibitem{Wolfenstein:1977ue}
L.~Wolfenstein,
``Neutrino Oscillations in Matter,''
Phys. Rev. D \textbf{17} (1978), 2369-2374.

\bibitem{Mikheyev:1985zog}
S.~P.~Mikheyev and A.~Y.~Smirnov,
``Resonance Amplification of Oscillations in Matter and Spectroscopy of Solar Neutrinos,''
Sov. J. Nucl. Phys. \textbf{42} (1985), 913-917.

\bibitem{Barger:1980tf}
V.~D.~Barger, K.~Whisnant, S.~Pakvasa and R.~J.~N.~Phillips,
``Matter Effects on Three-Neutrino Oscillations,''
Phys. Rev. D \textbf{22} (1980), 2718.

\bibitem{Xing:2000gg}
Z.~Z.~Xing,
``New formulation of matter effects on neutrino mixing and CP violation,''
Phys. Lett. B \textbf{487} (2000), 327-333
[arXiv:hep-ph/0002246 [hep-ph]].

\bibitem{Xing:2003ez}
Z.~Z.~Xing,
``Flavor mixing and CP violation of massive neutrinos,''
Int. J. Mod. Phys. A \textbf{19} (2004), 1-80
[arXiv:hep-ph/0307359 [hep-ph]].

\bibitem{Naumov:1991ju}
V.~A.~Naumov,
``Three neutrino oscillations in matter, CP violation and topological phases,''
Int. J. Mod. Phys. D \textbf{1} (1992), 379-399.

\bibitem{Mocioiu:2000st}
I.~Mocioiu and R.~Shrock,
``Matter effects on neutrino oscillations in long baseline experiments,''
Phys. Rev. D \textbf{62} (2000), 053017
[arXiv:hep-ph/0002149 [hep-ph]].

\bibitem{Xing:2016ymg}
Z.~Z.~Xing and J.~Y.~Zhu,
``Analytical approximations for matter effects on CP violation in the accelerator-based neutrino oscillations with E $\lesssim$ 1 GeV,''
JHEP \textbf{07} (2016), 011
[arXiv:1603.02002 [hep-ph]].






\end{thebibliography}
\end{document}